\RequirePackage{fix-cm}
\documentclass[smallextended]{svjour3}       
\smartqed  
\usepackage{graphicx}
\usepackage{bm}
\newcommand{\subscript}[1]{\ensuremath{_{\textrm{\footnotesize{#1}}}}}
\newcommand{\ket}[1]{\left| #1 \right\rangle}

\newcommand{\bra}[1]{\left\langle #1 \right|}
\newcommand {\gtrsim} {\ {\raise-.5ex\hbox{$\buildrel>\over\sim$}}\ }
 \journalname{Theoretical Chemistry Accounts}
\begin{document}

\title{Multireference X-Ray Emission and Absorption Spectroscopy calculations from Monte Carlo Configuration Interaction
}

\titlerunning{Multireference X-Ray Spectroscopy calculations from MCCI}        

\author{J. P. Coe         \and
        M. J. Paterson 
}


\institute{J. P. Coe \at
              Institute of Chemical Sciences, School of Engineering and Physical Sciences, Heriot-Watt University, Edinburgh, EH14 4AS, UK. 
           \and
            M. J. Paterson \at
             Institute of Chemical Sciences, School of Engineering and Physical Sciences, Heriot-Watt University, Edinburgh, EH14 4AS, UK. \\
                    \email{M.J.Paterson@hw.ac.uk}           
}

\date{Received: date / Accepted: date}

\maketitle

\begin{abstract}
We adapt the method of Monte Carlo configuration interaction to calculate core-hole states and use this for the computation of X-ray emission and absorption values. We consider CO, CH\subscript{4}, NH\subscript{3}, H\subscript{2}O, HF, HCN, CH\subscript{3}OH, CH\subscript{3}F, HCl and NO using a
6-311G** basis. We also look at carbon monoxide with a stretched geometry and discuss the dependence of its results on the cutoff used.  The Monte Carlo configuration interaction results are compared with EOM-CCSD values for X-ray emission and with experiment for X-ray absorption. Oscillator strengths are also computed and we quantify the multireference nature of the wavefunctions to suggest when approaches based on a single reference would be expected to be successful.
\end{abstract}

\section{Introduction}
\label{intro}

 X-ray absorption can be used to experimentally study core-electron excitations, e.g., as has been applied to small organic molecules in Ref.~\cite{1sXRAYexcitation}. While X-ray emission spectroscopy involves the initial ionization of a core electron followed by emission when the system adapts to remove the hole created in the core orbital.  This experimental method has recently facilitated the investigation of dynamics in water \cite{XrayDynamics}. More pertinent to this work it has previously been used to probe the energy of relaxation of a valence electron back to the core in, e.g, simple alcohols \cite{XrayEmissionAlcohols} and fluorine substituted methanes \cite{XrayEmissionFsubMethanes}.

A method based on damped coupled cluster response has been created \cite{CCresponseAbsorption} to calculate X-ray absorption values and, for CCSD response, agrees well with experiment for neon, carbon monoxide and water. This approach would be expected to work well when the ground state is not considered multireference.  Approaches based on density-functional theory (DFT) have also been created and shown to be accurate for small molecules, see, e.g., \cite{AgrenDFTabsorption,ChongDFTabsorption,OCDFT}. These DFT methods can be applied to larger systems, however the functional used will affect the accuracy and 
current functionals are considered to not cope well with multireference systems.

A successful computational approach to calculate the X-ray emission of many small molecules has been developed \cite{TDDFTandCCSDxray,CCSDxrayWater} using equation-of-motion coupled cluster singles and doubles (EOM-CCSD) \cite{EOMCCoverview}. However for EOM-CCSD to be accurate the initial state should be able to be described well by CCSD, i.e., it should have a clearly dominant configuration when treated exactly in a given basis and therefore not be considered multireference. In the method of Refs.~\cite{TDDFTandCCSDxray,CCSDxrayWater} a HF reference with a core hole is found using the maximum overlap method \cite{MOM} then this  is used for an EOM-CCSD calculation where the negative excitation energies are the emission values.
Such an approach allows multiple emission values to be accessed in a calculation however there may be problems with the convergence \cite{ConvergenceEOMCCSDcoreexcite} of the EOM-CCSD calculation and the approach becomes intractable beyond reasonably-sized molecules. Furthermore if the full configuration interaction (FCI) core-hole wavefunction is deemed multireference then EOM-CCSD would be expected to neglect the static correlation of the core-hole wavefunction and so may have difficulties with excitation energies.  Work on methods \cite{TDDFTandCCSDxray,TDDFTxray1} using time-dependent density-functional theory (TDDFT) offers the possibility of handling larger systems, but with a dependence on the approximations used and current functionals tend to have problems describing static correlation. For the further development of TDDFT approaches, in particular, the production of emission and absorption results for molecules of varying multireference character would therefore be useful. These data could also be used in improving the parameters in spin-component scaled configuration interaction with single substitutions and perturbative doubles SCS-CIS(D) \cite{scsCISD} which also offers the possibility of emission calculations for larger molecules.

Here we consider a complementary approach that is also limited to molecules that are not too large but should be able to deal with multireference situations and is not affected by convergence issues for a single emission calculation. To do this we adapt the method of Monte Carlo configuration interaction (MCCI) \cite{mcciGreer98,mccicodeGreer} to describe core-hole wavefunctions.  MCCI stochastically builds up a wavefunction with the aim of capturing many of the important aspects of the FCI wavefunction by accounting for both static and dynamic correlation to some degree, but using only a very small fraction of configurations. The method has been successfully applied to single point energies \cite{MCCIGreer95}, dissociation energies \cite{dissociationGreer,MCCIfirstrowdiss}, electronic excitations \cite{GreerMcciSpectra,2013saMCCI}, ground-state \cite{MCCIpotentials,MCCImetaldimers} and excited potential curves \cite{2013saMCCI}, multipole moments \cite{MCCIdipoles} and higher-order dipole properties up to the second hyperpolarizability \cite{MCCIhyper}.

We calculate the X-ray emission energies at equilibrium geometries for CO, CH\subscript{4}, NH\subscript{3}, H\subscript{2}O, HF, HCN, CH\subscript{3}OH, CH\subscript{3}F, HCl and NO. The emission energy for CO at a stretched geometry of $R=4$ $a_{0}$~  is also considered and we also look at the X-ray absorptions for the same set of molecules. We compare the emission energies with EOM-CCSD results of Ref.~\cite{TDDFTandCCSDxray} when possible. These EOM-CCSD results have very good agreement with the available experimental studies. The absorption energies are compared in relation to  available experimental results in the literature. The oscillator strength and multireference character of the states of interest are also computed and discussed.  We note that this MCCI approach offers the possibility of multireference computational results for emission and absorption.  We are not attempting to offer an improvement over EOM-CCSD for all systems and acknowledge that EOM-CCSD will be more accurate for systems that do not have significant multireference character.  However we hope that the results in this work will encourage tests, and possibly improvements, of EOM coupled cluster and TDDFT emission calculations on more challenging multireference systems such as stretched geometries, nitric oxide and the carbon dimer.
\section{Methods}
\label{sec:methods}
MCCI \cite{mcciGreer98,mccicodeGreer} randomly augments the configuration space by making single and double substitutions in the current selection of configuration state functions (CSFs) so that symmetry is preserved. By using configuration state functions the MCCI wavefunction is guaranteed to be a spin eigenfunction. The Hamiltonian matrix is then constructed using these configurations and diagonalized. Any newly added configurations with an absolute coefficient, subject to appropriate normalization \cite{GreerMcciSpectra}, less than the cutoff (c\subscript{min}) are discarded and every ten iterations all configurations falling into this category are removed. After sixty iterations, the process continues until convergence in the energy, as described in Ref.~\cite{GreerMcciSpectra}, is observed to $0.001$ Hartree. The usual starting point is the configuration formed from the occupied Hartree-Fock molecular orbitals. The molecular orbitals and their required integrals are calculated using COLUMBUS \cite{COLUMBUS7}. 

Core-hole states could be calculated in MCCI by considering very high energy eigenvalues however for a stable calculation this would be expected to require all lower eigenvalues and so would not be feasible. Hence we extend the method to ground-state calculation restricted to a single occupied core orbital.

 For X-ray emission results, we initially perform a standard MCCI calculation on the cation of the required symmetry.  We then use MCCI to calculate the energy of the cation when the orbital containing the core hole is restricted to be singly occupied in all configurations.  This is achieved by starting with a reference where the core orbital of interest is singly occupied then only allowing substitutions that preserve this. One subtlety is that MCCI employs CSFs and uses the genealogical scheme \cite{mcciGreer98} to ensure all orbital lists correspond to linearly independent CSFs. This means that the frozen single occupied orbital may be alpha spin in some lists and beta spin in others.  As only non-frozen orbitals are available for substitution into existing configurations then for a randomly chosen configuration we check which spin does not have the frozen single occupied orbital and then allow the possibility of all but the double occupied frozen orbitals to be replaced in this spin.   

To calculate an X-ray absorption energy we begin with the neutral molecule then repeat the calculation for a core-hole state of the required symmetry with the lowest energy orbital singly occupied in all configurations. 

 Below we summarize the use of MCCI for core-hole states starting with a reference consisting of a single occupied core Hartree-Fock molecular orbital.
\begin{enumerate}
\item{Create new configurations by random single and double substitutions in the current set of configurations so that symmetry and the frozen orbitals are preserved.}
\item{Create the Hamiltonian and overlap matrices then diagonalize.}
\item{Any new configurations with absolute coefficient less than c\subscript{min} are removed.}
\item{Every ten iterations all configurations are considered as candidates for deletion.}
\item{The procedure is repeated until the energy has converged.}
\end{enumerate}

To calculate oscillator strengths between the two states of interest the following equation is employed
\begin{equation}
f_{ab}=\frac{2}{3} \Delta E | \bm{D}_{ab}|^{2}.
\end{equation}
Here
\begin{equation}
 \bm{D}_{ab}=\bra{\Psi_{a}} \bm{\hat{r} }\ket{\Psi_{b}}.
\end{equation}

We approximately quantify the multireference nature of the MCCI wavefunctions by using the approach introduced in Ref.~\cite{MCCImetaldimers}.  There 
\begin{equation}
MR=\sum_{i} |c_{i}|^{2}-|c_{i}|^{4}
\end{equation}
 is calculated with an approximate normalization for configuration state functions such that $\sum |c_{i}|^{2}=1$.  Here a value of zero signifies that the wavefunction is single reference and one is approached as the system becomes more multireference. Previous work \cite{MCCImetaldimers} saw that the MR of an MCCI wavefunction for the strongly multireference chromium dimer when using a cc-pVTZ basis ranged from around $0.8$ to almost $1$ as the bond length was varied. In Ref.~\cite{MCCIhyper} the value for HF in a aug-cc-pVDZ basis was found to be $0.30$ for an MCCI wavefunction suggesting that this system is amenable to being modelled using methods based on a single reference.

\section{Results}
\label{sec:results}
\subsection{Emission Energies}
We first model the X-ray emission energy following the ionization of an electron from the lowest lying core orbital. If symmetry is used then both states are completely symmetric unless otherwise noted.  We compare MCCI values using c\subscript{min}$=5\times 10^{-4}$ with the results of Ref.~\cite{TDDFTandCCSDxray} for molecules that contain first row atoms and with one example (HCl) of a molecule containing a second row atom.  We use the experimental geometry of the neutral molecule throughout except for methanol and CH\subscript{3}F where we optimize the geometry when using MP2 with cc-pVTZ.  The calculations for CH\subscript{4}, NH\subscript{3}, H\subscript{2}O and HF used one frozen orbital while the other calculations used two except for HCl where five were employed.

\begin{table}[h]
\caption{MCCI emission energies and oscillator strengths at c\subscript{min}$=5\times 10^{-4}$ with a 6-311G** basis when using the lowest-lying core hole in the ionized molecule compared with experimental and EOM-CCSD results as listed in Ref.~\cite{TDDFTandCCSDxray}.} \label{tbl:MCCIcoreholeion6311}
\begin{tabular}{lllll}
\hline\noalign{\smallskip}
  Molecule & MCCI (eV) & f & Exp (eV) &   EOM-CCSD (eV)  \\
\noalign{\smallskip}\hline\noalign{\smallskip}
CO  & $529.7$  & $7.1\times 10^{-3}$  & -  & 526.6 \\
CH\subscript{4} $A_{1}\rightarrow B_{2}$   & $277.3$  & $3.0\times10^{-2}$ & 276.3 & 276.2  \\
NH\subscript{3}   & $395.8$  & $4.3\times10^{-2}$ & 395.1 & 395.0 \\
H\subscript{2}O   & $525.8$  & $4.3\times10^{-2}$ & 525.1 & 525.4  \\
HF   & $674.7$  & $4.5\times10^{-2}$ & - & 674.5  \\
HCN   & $393.9$  & $3.3\times10^{-2}$ & - & 393.1 \\
CH\subscript{3}OH   & $529.7$  & $4.7\times10^{-2}$ & 523.9 & 522.2  \\
CH\subscript{3}F   & $681.3$  & $3.0\times10^{-2}$ & 675.6 & 675.5 \\
CO ($R=4$ $a_{0}$)  & $528.0$  & $1.5\times 10^{-2}$ & - & - \\
HCl            & $2821.0$       &  $4.6\times 10^{-3}$ & -  & 2811.6   \\
NO            & $544.0$       &  $9.1\times 10^{-6}$ & -  & -   \\
C\subscript{2} ($R=1.25$ \AA) & $287.8$ & $0$ & -  & -\\
\noalign{\smallskip}\hline
\end{tabular}
\end{table}

 The results for emission energies are displayed in table~\ref{tbl:MCCIcoreholeion6311} and, except for CH\subscript{3}OH, CH\subscript{3}F and CO, are close to, but slightly higher than, the available EOM-CCSD results of Ref.~\cite{TDDFTandCCSDxray} which themselves are in excellent agreement with experiment where available. This suggests that the core-hole state may not be described quite as well in this MCCI procedure as the cation.  We note that these values for oscillator strengths are of a similar order of magnitude to those calculated with EOM-CCSD and a u6-311G** basis in Ref.~\cite{TDDFTandCCSDxray} while HCN and CH\subscript{3}F used configuration interaction singles. The oscillator strengths demonstrate that the transitions are not forbidden within the dipole approximation except perhaps for NO.  However for nitric oxide with a final state of $B_{1}$ symmetry we find that $f=2.0\times10^{-2}$ and the emission energy is $535.2$ eV while for a core hole in the second lowest orbital the MCCI emission value is $403.61$ eV ($f=4.0\times10^{-4}$) which is in good agreement with the experimental result \cite{NOemission} of X-ray lines around $403$ to $402$ eV assigned to a core hole in N 1s. For the latter core-hole state we find that $MR=0.87$.  This is an important result as it demonstrates that this approach can give good agreement with experiment when the multireference nature is very high. We also find that for the carbon dimer with a bond length of $1.25$ angstrom that the multireference nature for the core-hole state is high at $MR=0.76$ and the emission is $287.8$eV. For the non-forbidden $A_{g} \rightarrow B_{2u}$ transition we calculate $f=4.7\times10^{-2}$ and an emission energy of $289.4$ eV. 

In table~\ref{tbl:MCCIcoreholeion6311percenterror} we display the percentage error with the EOM-CCSD results.  We see that there is very close agreement with the EOM-CCSD results, with the largest difference for methanol at $1.4\%$.  HCl was considered in Ref.~\cite{TDDFTandCCSDxray} using u6-311G**  so we cannot compare directly and an experimental result is not available to our knowledge, but we note that their value was $2811.6$ eV and that our result compared with this has an error of $0.3\%$.  When using the cc-pCVDZ  basis we calculate the emission as $2821.0$ eV while the EOM-CCSD result \cite{TDDFTandCCSDxray} was $2805.9$. A Hartree-Fock calculation with the Douglas-Kroll-Hess Hamiltonian in MOLPRO \cite{MOLPRO_brief2012} suggests that in this basis the energy of the lowest energy Hartree-Fock orbital is reduced by $8.1$ eV. This allows us to estimate the MCCI value when corrected for relativistic effects as $2829.1$ eV. For the cc-pCVTZ basis the MCCI result is $2819.6$ eV and the approximate correction for relativistic effects gives $2827.7$ eV.

As MCCI uses a random process to choose configurations we check that the results are sufficiently robust at this cutoff by repeating the calculations for water a total of ten times.  We find that the mean emission energy is in agreement to one decimal place with the single calculation of table~\ref{tbl:MCCIcoreholeion6311} at $525.8$ eV with a standard error of $0.0005$ eV.

\begin{table}[h]
\caption{Percentage differences when compared with EOM-CCSD \cite{TDDFTandCCSDxray} when using MCCI at c\subscript{min}$=5\times 10^{-4}$ with a 6-311G** basis when considering the lowest-lying core hole in the ionized molecule.} \label{tbl:MCCIcoreholeion6311percenterror}
\begin{tabular}{ll}
\hline\noalign{\smallskip}
  Molecule & Percentage Difference  \\
\noalign{\smallskip}\hline\noalign{\smallskip}
CO  & $0.6\%$    \\
CH\subscript{4}   & $0.4\%$   \\
NH\subscript{3}   & $0.2\%$   \\
H\subscript{2}O   & $0.08\%$    \\
HF   & $0.04\%$    \\
HCN   & $0.2\%$    \\
CH\subscript{3}OH   & $1.4\%$    \\
CH\subscript{3}F   & $0.9\%$    \\
\noalign{\smallskip}\hline
\end{tabular}
\end{table}

In table~\ref{tbl:MCCIcoreholeMR} we display the multireference values for the MCCI wavefunctions.  When neither the cation nor the core-hole state is deemed multireference we display the molecular orbital transition when considering the most significant configuration ($|c| \gtrsim 0.9$) in each state.  For the systems that we compare with the EOM-CCSD values at the neutral equilibrium geometry the core-hole cation would not be considered multireference except for perhaps carbon monoxide. This suggests that the use of EOM-CCSD is indeed appropriate for these systems and even for carbon monoxide we note that the percentage difference is only $0.6\%$ (table~\ref{tbl:MCCIcoreholeion6311percenterror}) although in Ref.~\cite{NOemission} there were two experimental emission values for carbon monoxide assigned to sigma orbitals at 522.3 eV and 530.2eV, the EOM-CCSD result at 525.6eV lies between these values while MCCI is close to the higher value, however this was noted as being a very weak line.  We note that the core-hole state is deemed multireference for NO suggesting that methods based on a single reference could perform poorly in this case.

\begin{figure}[ht]
\includegraphics[width=.6\textwidth]{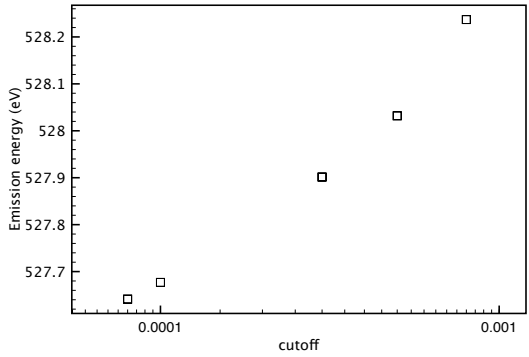}
\caption{MCCI results for the emission energy of carbon monoxide ($R=4$ $a_{0}$) against c\subscript{min} on a logarithmic scale when using the 6-311G** basis.}\label{fig:COvCmin}
\end{figure}

\begin{figure}[ht]
\includegraphics[width=.6\textwidth]{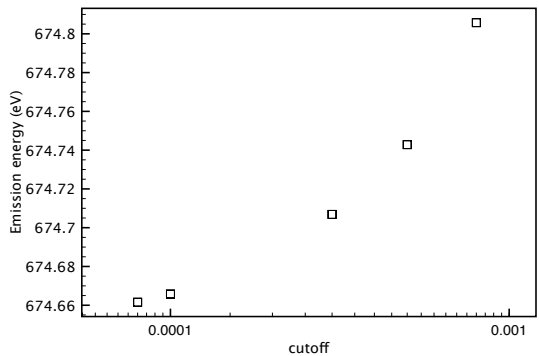}
\caption{MCCI results for the emission energy of hydrogen fluoride against c\subscript{min} on a logarithmic scale when using the 6-311G** basis.}\label{fig:HFvCmin}
\end{figure}

For carbon monoxide at a stretched geometry, both considered MCCI states are strongly multireference.  We note that this stretched geometry results in a $1.7$ eV change in the emission energy (table~\ref{tbl:MCCIcoreholeion6311}).  We investigate the effect of varying c\subscript{min} on the emission value for the stretched molecule. The FCI space is around $10^{9}$ Slater determinants when symmetry is included while for the lowest cutoff considered we required 73883 CSFs for the cation and 115035 when the core hole is used. For c\subscript{min}$=5\times10^{-4}$ 8702 and 10949 CSFs respectively were required.  In Fig.~\ref{fig:COvCmin} we see although the emission energy is non-variational it lowers with cutoff for the points considered. The plot suggests that for this challenging multireference system the results are still a little away from full convergence with respect to cutoff but we would not expect the emission energy to drop below around $527.6$ eV. The emission energy reduces by around $0.35$ eV on lowering c\subscript{min} from the $5\times10^{-4}$ value used for calculations in this paper to $1\times10^{-4}$ and then by $0.04$ eV to $527.64$ eV on further reduction of the cutoff to $8\times10^{-5}$.  In Fig.~\ref{fig:HFvCmin} we see that for a system with low multireference character, hydrogen fluoride, there is again a decrease with cutoff but here the results seem much closer to convergence: the emission energy only reduces by $0.08$ eV on lowering c\subscript{min} from $5\times10^{-4}$ to $1\times10^{-4}$ and then by $0.004$ eV on reducing c\subscript{min} to $8\times10^{-5}$.

\begin{table}[h]
\caption{MCCI multireference character at c\subscript{min}$=5\times 10^{-4}$ with a 6-311G** basis for the ionized molecule with and without a hole in the lowest lying core orbital.} \label{tbl:MCCIcoreholeMR}
\begin{tabular}{lll}
\hline\noalign{\smallskip}
  Molecule & Cation with core-hole MR  & Cation MR   \\
\noalign{\smallskip}\hline\noalign{\smallskip}
CO  & $0.65$  & $0.35$    \\
CH\subscript{4} $1b_{2}\rightarrow 1a_{1}$  & $0.43$ & $0.34$    \\
NH\subscript{3} $5a\rightarrow 1a$  & $0.33$  & $0.18$  \\
H\subscript{2}O $3a_{1}\rightarrow 1a_{1}$    & $0.29$ & $0.15$    \\
HF $3a_{1}\rightarrow 1a_{1}$    & $0.23$  & $0.11$   \\
HCN   & $0.45$  & $0.64$   \\
CH\subscript{3}OH    & $0.36$ & $0.75$   \\
CH\subscript{3}F $9a\rightarrow 1a$   & $0.43$ & $0.33$    \\
CO ($R=4$ $a_{0}$)  & $0.80$ & $0.87$     \\
HCl      $5a_{1}\rightarrow 1a_{1}$       &  $0.35$ & $0.20$            \\
NO             &  $0.88$  & $0.56$          \\
C\subscript{2} ($R=1.25$ \AA) & $0.76$ & $0.72$ \\
\noalign{\smallskip}\hline
\end{tabular}
\end{table}

Table~\ref{tbl:ExperimentalErrors} displays the percentage errors of our MCCI calculations and the EOM-CCSD calculations of Ref.~\cite{TDDFTandCCSDxray} with the available experimental values listed in Ref.~\cite{TDDFTandCCSDxray}.  We see that EOM-CCSD is closer to the experimental results in the considered cases, however this is expected as this selection of molecules are not considered to have substantial multireference character for the core-hole state (table~\ref{tbl:MCCIcoreholeMR}). 

\begin{table}[h]
\caption{ Percentage differences when compared with experiment when using MCCI at c\subscript{min}$=5\times 10^{-4}$ with a 6-311G** basis and for the EOM-CCSD results of Ref.~\cite{TDDFTandCCSDxray}.} \label{tbl:ExperimentalErrors}
\begin{tabular}{lll}
\hline\noalign{\smallskip}
  Molecule & MCCI & EOM-CCSD \\
\noalign{\smallskip}\hline\noalign{\smallskip}
CH\subscript{4}   & $0.35\%$ & $0.04\%$   \\
NH\subscript{3}   & $0.18\%$  & $0.03\%$   \\
H\subscript{2}O   & $0.13\%$ & $0.06\%$     \\
CH\subscript{3}OH   & $1.10\%$  & $0.32\%$    \\
CH\subscript{3}F   &  $0.84\%$ & $0.01\%$  \\
\noalign{\smallskip}\hline
\end{tabular}
\end{table}

We use three systems as an example of the computational cost when using twelve processors for emission with increasing multireference character of the core-hole state. For HF the cation requires around 1 minute and 1320 CSFs while
the core-hole state used around 5 minutes and 2724 CSFs. The CO cation required 13 minutes and 5838 CSFs and the core-hole state uses 53 minutes and 11316 CSFs. For CO with a stretched geometry 55 minutes and 8702 CSFs are needed for the cation
while the core-hole state needed 1 hour and 38 minutes and 10949 CSFs. We note that in all three considered cases the core-hole state is more challenging to compute and the cost increases with the multireference character.  

\subsection{Absorption Energies}

We now consider X-ray excitation energies of an electron from the lowest lying core orbital in the same range of molecules rather than the emission energy. The results are presented in table~\ref{tbl:MCCIcoreholeneutral6311}. Unless otherwise stated, when symmetry is used we consider transitions between states classed as totally symmetric. For the $A_{1} \rightarrow A_{1}$ transition in CH\subscript{4}, we find $289.0$ eV compared with $287.1$ eV for the experimental result \cite{1sXRAYexcitation}.  The result stands out as the f value of $6\times 10^{-11}$ indicates that this transition is forbidden within the dipole approximation. Hence we also calculate a core-hole molecule of B\subscript{2} symmetry. This gives $290.4$ eV and $f=3\times 10^{-2}$ while using the first $A_{1}$ excited core-hole state gives $290.8$ and $f=3\times 10^{-2}$. We note that the experimental results range from $288$ eV to $290$ eV for transitions assigned as to a t\subscript{2} orbital \cite{1sXRAYexcitation}.  

For water, the experimental absorption \cite{PhysRevA.47.1136} for the first transition assigned to an $A_{1}$ state is $534.0$ eV which is also close to the MCCI calculation.  Absorption spectra for methanol have been calculated in Ref.~\cite{MethanolAbsorption} with the first peak at around $534$ eV and a stronger absorption at about $537$ eV which the MCCI result is close to. For CO the largest photoionization yield is between $534$ and $535$ eV in Ref.~\cite{COabsorption}.  The MCCI value is somewhat higher but this for an excitation of the same symmetry not an excitation to a $\pi$ orbital. For the excitation to $B_{1}$ symmetry we find much better agreement as the absorption energy is $535.6$ eV with $f=3.5\times10^{-2}$. For Nitric oxide Ref.~\cite{COabsorption} finds experimentally that absorption requires between around $532$ eV and $534$ eV for excitation to $^{2}\Sigma^{-}$ or $^{2}\Delta$. These states are of $A_{2}$ symmetry when using $C_{2v}$ therefore agreeing with our result. For the $B_{2} \rightarrow B_{2}$ transition we found $f=2.9\times 10^{-5}$ with an absorption energy of $542.9$ eV.

The damped coupled cluster linear response results of Ref.~\cite{CCresponseAbsorption} for water and carbon monoxide are also in agreement with experiment with the exception of those from coupled cluster singles which are too high.  The CCSD-NR result for water is 535.68 eV while for the CO excitation to a $\pi$ orbital it is 535.85 eV.  For water and carbon monoxide the multireference
character is not high for the molecule (table~\ref{tbl:MCCIcoreholeNeutralMR}) suggesting that this approach would be expected to be effective.  For these absorption results, methods based on DFT with large basis sets have found 533.89 eV for water \cite{ChongDFTabsorption} while for CO the energies were 534.21 eV \cite{ChongDFTabsorption}, 533.0 eV \cite{OCDFT} and, depending on the functional, 535.1 eV to 536.1 eV  \cite{AgrenDFTabsorption}.  

\begin{table}[h]
\caption{MCCI X-ray absorption energies and oscillator strengths at c\subscript{min}$=5\times 10^{-4}$ with a 6-311G** basis for the lowest energy core hole in the neutral molecule  compared with experimental results \cite{PhysRevA.47.1136,MethanolAbsorption,COabsorption,HCNabsorption,HClabsorption}. } \label{tbl:MCCIcoreholeneutral6311}
\begin{tabular}{llll}
\hline\noalign{\smallskip}
  Molecule & MCCI (eV) & f &  Exp (eV) \\
\noalign{\smallskip}\hline\noalign{\smallskip}
CO  & $544.5$  & $1.4\times 10^{-3}$ & 538.9  \\
CH\subscript{4}     & $289.0$  & $5.9\times10^{-11}$ &  -   \\
NH\subscript{3}   & $402.8$  & $6.0\times10^{-3}$ & -  \\
H\subscript{2}O   & $535.7$  & $1.2\times10^{-2}$  & 534.0 \\
HF   & $688.7$  & $1.9\times10^{-2}$ & - \\
HCN   & $405.2$  & $3.6\times10^{-3}$ & 401.8 \\
CH\subscript{3}OH   & $538.8$  & $9.3\times10^{-3}$ & 534.1  \\
CH\subscript{3}F   & $693.2$  & $1.4\times10^{-3}$  & - \\
CO ($R=4$ $a_{0}$)  & $531.2$  & $3.0\times 10^{-2}$ & -  \\
HCl            & $2832.0$       &  $3.9\times 10^{-3}$ & 2823.9    \\
NO   ($B_{2} \rightarrow A_{2}$)          & $534.1$       &  $3.2\times 10^{-2}$ & $\sim532$ to $\sim534$     \\
\noalign{\smallskip}\hline
\end{tabular}
\end{table}

Table~\ref{tbl:MCCIcoreholeNeutralMR} displays how all of the core-hole states for the neutral molecules appear to have multireference character. This continues in the core-hole state of B\subscript{1} symmetry for carbon monoxide ($MR=0.71$).
The core-hole state also exhibits multireference character for the $B_{2}$ symmetry methane  result $MR=0.66$. Therefore a single calculation approach based on a single reference would be expected to encounter difficulties. However earlier work \cite{CH4uhf} with unrestricted HF wavefunctions also achieved accurate results for CH\subscript{4} and remarked that the neglect of correlation cancels out to a large extent in this case when using the difference in energy between the neutral molecule with and without a core hole.  

\begin{table}[h]
\caption{MCCI multireference character at c\subscript{min}$=5\times 10^{-4}$ with a 6-311G** basis for the neutral molecule with and without a hole in the lowest lying core orbital.} \label{tbl:MCCIcoreholeNeutralMR}
\begin{tabular}{lll}
\hline\noalign{\smallskip}
  Molecule & MR & core-hole MR   \\
\noalign{\smallskip}\hline\noalign{\smallskip}
CO  & $0.47$  & $0.81$   \\
CH\subscript{4}   & $0.37$  & $0.77$  \\
NH\subscript{3}   & $0.31$  & $0.74$  \\
H\subscript{2}O   & $0.32$  & $0.73$  \\
HF   & $0.26$  & $0.71$  \\
HCN   & $0.51$  & $0.86$  \\
CH\subscript{3}OH   & $0.45$  & $0.76$  \\
CH\subscript{3}F   & $0.41$  & $0.76$  \\
CO ($R=4$ $a_{0}$)  & $0.89$  & $0.89$   \\
HCl            & $0.33$       &  $0.78$     \\
NO              & $0.52$       &  $0.75$     \\
\noalign{\smallskip}\hline
\end{tabular}
\end{table}

As examples of the computational cost for absorption when using twelve processors we consider three systems of increasing multireference character.  For HF the calculation for the molecule needed around 1 minute and used 1339 CSFs while the core-hole state required 13 minutes and used 4164 CSFs. CO needed 5 minutes and 4576 CSFs for the molecule. The core-hole state needed 1 hour and 14 minutes and 10430 CSFs. For CO with a stretched geometry the molecule needed 45 minutes and 8836 CSFs while the core-hole state required 1 hour 47 minutes and 9639 CSFs. Similarly to the emission calculations the computational cost increased with multireference character and the core-hole states were more challenging.
\section{Summary}
\label{sec:summary}
We put forward a complementary approach to calculate X-ray emission and absorption energies for reasonably-sized molecules using Monte Carlo configuration interaction (MCCI).  This method should be able to cope sufficiently well whether the system is deemed to be well described by methods based on a single-reference or if multireference approaches are required. 

We saw that at equilibrium geometries the X-ray emission energies had very small percentage differences with the available EOM-CCSD results of Ref.~\cite{TDDFTandCCSDxray}. When we quantified the multireference nature of the MCCI wavefunction we observed that the results suggested that the core-hole wavefunction tended not to be multireference in character and so EOM-CCSD would be expected to work very well for most of the systems.  Nitric oxide was one of the exceptions to this where its core-hole state was deemed to be strongly multireference and the MCCI result for emission following the creation of a hole in the second lowest energy orbital compared  well with experiment. This suggests that similar open shell systems may pose difficulties for emission calculations when using approaches built around a single reference. We also considered carbon monoxide at a stretched geometry and saw that the system would be considered multireference with an accompanying change in the X-ray emission of $1.7$ eV. 

We also looked at the X-ray absorption of the molecules and compared the MCCI results with experimental data when available. For methane we found reasonably good agreement with experiment for the excitation of an electron from the lowest lying core orbital.  The results with water, ethanol, hydrogen cyanide and nitric oxide also fitted in with known experimental values. The value for hydrogen chloride was about 8 eV higher than experiment.  The largest absorption energy in carbon monoxide was higher than experiment but for excitation to $B_{1}$ symmetry we found much better
agreement with experiment.  Interestingly the multireference character of the core-hole MCCI wavefunction was fairly large implying that methods based around the unrelaxed core-hole single-reference may encounter difficulties for these absorption calculations. 

This approach can be straightforwardly extended to consider holes in orbitals that are not the lowest in energy and we have illustrated this on nitric oxide. When each wavefunction has only one significant configuration then we can label the transition using two molecular orbitals, however  when dealing with multiconfigurational wavefunctions although we choose the core-hole orbital, it is not trivial, or perhaps possible, to label the transition in terms of a single excitation using molecular orbitals. The use of natural transition orbitals \cite{Martin2003}, as used for wavepackets created by X-rays \cite{wavepacketNatTranOrbs}, or natural transition geminals \cite{NatGeminals} may allow this transition to be assigned a compact description.

These calculations of X-ray emission and absorption for reasonably sized molecules with strong multireference character should provide useful data for improving approximations in methods for larger systems such as time-dependent density-functional theory.

\begin{acknowledgements}
We thank the European Research Council (ERC) for funding under the European Union's Seventh Framework Programme (FP7/2007-2013)/ERC Grant No. 258990. 
\end{acknowledgements}

\providecommand{\noopsort}[1]{}\providecommand{\singleletter}[1]{#1}%

\end{document}